\pdfoutput=1
\documentclass[conference]{IEEEtran}
\usepackage[normalem]{ulem}
\usepackage{amsmath}
\usepackage{algorithm}
\usepackage[noend]{algpseudocode}
\usepackage{xcolor}

\ifCLASSINFOpdf
   \usepackage[pdftex]{graphicx}
   \graphicspath{{../pdf/}{../jpeg/}}
   \DeclareGraphicsExtensions{.pdf,.jpeg,.png}
\else
   \usepackage[dvips]{graphicx}
   \graphicspath{{../eps/}}
   \DeclareGraphicsExtensions{.eps}
\fi
\usepackage[font=footnotesize]{subfig}

\usepackage{url}

\begin{document}
%
\title{The Effect of Instruction Padding on SFI Overhead}

\author{\IEEEauthorblockN{Navid Emamdoost}
\IEEEauthorblockA{University of Minnesota\\
navid@cs.umn.edu}
\and
\IEEEauthorblockN{Stephen McCamant}
\IEEEauthorblockA{University of Minnesota\\
mccamant@cs.umn.edu}
}


%


\maketitle

\begin{abstract}
Software-based fault isolation (SFI) is a technique to isolate a
potentially faulty or malicious software module from the rest of a
  system using instruction-level rewriting.
  SFI implementations on CISC architectures, including Google Native
  Client, use instruction padding to enforce an address layout
  invariant and restrict control flow.
  However this padding decreases code density and imposes runtime
  overhead.
We analyze this overhead, and show that it can be reduced by allowing
some execution of overlapping instructions, as
long as those overlapping instructions are still safe according to the
original per-instruction policy.
  We implemented this change for both 32-bit and 64-bit
x86 versions of Native Client, and analyzed why the performance benefit is higher on 32-bit.
The optimization leads to a consistent decrease in the number of
  instructions executed and savings averaging 8.6\%
  in execution time (over compatible benchmarks from SPECint2006)
  for x86-32.
  We describe how to modify the validation algorithm to check the more
  permissive policy, and extend a machine-checked Coq proof to confirm
  that the system's security is preserved.
\end{abstract}


%

\section{Introduction}

In application development it is desirable to have extensibility via integrating multiple (and possibly separately developed) software modules. 
This lets an application achieve new functionality or modify existing functionality with the least change in its existing code. 
Browser plug-ins are good examples of this approach. Although  extensions give flexibility for application development, protecting 
application security is crucial, too. A buggy or malicious extension can subvert the application. Software-based Fault Isolation (\textit{SFI}) is an 
effective solution that constrains an untrusted module by loading and executing it with  separated code and data segments from the hosting application. 
SFI enforces security policies which ensure that the untrusted code can only  access its own resources and can only  invoke a list of pre-approved system calls.

Native Client (\textit{NaCl} for short)~\cite{yeeNaCl} is a Google Chrome browser service that allows incorporating native (e.g., x86) code into a web application. 
This gives the web application the computational performance of native code. To protect the browser's security from  potentially buggy or malicious native code, 
NaCl enforces a sandboxing policy. This policy ensures the native code  executes just within its own code, accesses just its own data, and communicates with 
the browser just through predefined interfaces.  To be NaCl-compatible, native code  needs to be compiled with the NaCl build chain. The NaCl compiler inserts 
checking instructions and aligns instructions to enforce sandboxing policies. A validator is responsible for checking policy adherence before the 
native code may execute.

NaCl implements SFI for both CISC  (x86-32 and x64) and RISC (ARM and MIPS) architectures. This gives the browser native-code performance across 
the most common underlying architectures. In the CISC implementation, NaCl incorporates a conservative instruction padding scheme like the one first 
proposed in PittSFIeld~\cite{mccamantPittS}. While this padding scheme is crucial to enforce the sandboxing policy (via instruction alignment), it imposes a 
runtime overhead on the execution by introducing more instructions to be executed and reducing the effectiveness of instruction caching and prefetching.

In this paper we improve performance by changing the NaCl padding scheme. In addition to implementing our changes in the instruction alignment of the 
NaCl build chain, we changed the validator appropriately to ensure our changes do not disrupt NaCl sandboxing policy enforcement.
We implemented and evaluated our padding scheme for both x86-32 and
x64 NaCl, the two NaCl architectures that use artificial alignment.
We chose NaCl for our evaluation because it is well-known and
represents the state of the art, but a similar change would apply to
other systems that use artificial alignment to limit control flow.

In May 2017, Google announced that it is phasing out support for NaCl
in most parts of Google Chrome/Chromium in favor of WebAssembly, which
has better cross-browser support~\cite{nelsonChromiumBlog}.
This is an understandable choice because Chrome already has a JavaScript
JIT in its trusted computing base (TCB), but the static compilation
and small verification TCB offered by SFI are appealing in other
contexts as well.
For instance, NaCl has also been applied to provide isolation for
distributed analytics~\cite{Rad2014} and edge
clouds~\cite{JonathanROCW2017}, and for native libraries included in
the JDK~\cite{SiefersTM2010} or used in Android
apps~\cite{AthanasopoulosK2016}.

We proved the correctness of the changed validator using the Coq proof assistant~\cite{Coq}. For this project, we incorporated a greedy algorithm for 
padding instructions. Our evaluations show that changing the  padding scheme can improve code performance, though in some cases the analysis is 
complicated by other CPU optimizations like branch prediction. To support performance analysis we adapted the Valgrind/Cachegrind tool to monitor Native Client execution.

The rest of this paper is organized as follows: in Section~\ref{background} we discuss some background notions essential to the rest of the paper. 
In Section~\ref{related} we survey previous projects related to  SFI. In Section~\ref{cbi-nacl} we discuss our changes to the NaCl build chain and validator. 
Section~\ref{eval} describes our evaluation method and presents the results and their analysis. Section~\ref{challenge} talks about some technical obstacles 
we faced during this project and the ways we solved them; and Sections \ref{future} and \ref{concl} present the future work and the conclusion, respectively.

\section{Background}
\label{background}
In this section we go through some related background. First we introduce the notion of Software-based Fault Isolation (SFI) and its advantages. Then we compare this  to another  closely related security mechanism, and discuss branch prediction in a CPU and its relevance to our project.
\subsection{Software-Based Fault Isolation}
In the context of software systems, fault isolation is the ability to contain a potentially faulty module from other parts of the system, meaning that untrusted module failure does not affect other modules. In a more extreme case, even a malicious untrusted module should not be able to interfere with other modules' functionality. One approach to this problem is loading each module into a separate address space. This can be supported  by hardware, and guarantee there is no way to access other modules' resources, e.g. OS process isolation.

Such hardware-based  isolation is robust but inefficient as inter-module communication incurs context switch overhead. 
Wahbe et al.~\cite{wahbeSFI}  propose a software-based solution. In this approach isolation is provided within a single hardware address space. 
The untrusted module has its own code and data region (the sandbox), and is prohibited from jumping or accessing other modules' regions.
This prohibition is implemented by rewriting the untrusted module code to limit  control transfer or data access instructions. In case of direct addressing (like  \texttt{jmp  0x100019b}), such instructions can be checked statically even before the execution starts. That is because knowing the destination address enables the check (requiring non-writable code during execution guarantees the checked address will not change at runtime).

The key challenge is with indirect addressing (instructions like \texttt{jmp~*\%ecx} or \texttt{mov \$0x63,(\%ecx)}); in these cases the destination address cannot be resolved until the point of instruction execution. Therefore SFI inserts some checking instructions to ensure the register contains a valid address. The challenge is that these checks must be efficient and unable to be bypassed. In Section~\ref{related} we will go through different SFI techniques that address this challenge.

A related security mechanism  is Control Flow Integrity (CFI) which first was proposed  by Abadi et al.~\cite{abadiCFI}. CFI tries to enforce \textit{control flow graph  (CFG) integrity} by restricting all indirect control transfers to go only to intended targets. In the original approach of Abadi et al., a system first statically determines valid targets for call and return instructions. To enforce the extracted CFG, an ID is used for valid targets and is placed right before the target location. Then each indirect call and return instruction is instrumented to first check the target address ID and to jump only if the ID is correct. 
 This technique also requires non-writable code because IDs must be tamper-proof and non-reproducible.

Compared to SFI, CFI enforces a smaller set of places an indirect jump can target. In SFI it is sufficient that no indirect control flow evades the sandboxed code region, but in CFI it is further required that the address be an intended target address. SFI is sometimes described as being based on a weak form of CFI, where the set of valid targets is all the sandboxed area. However the control-flow enforcement used in SFI would not be a strong CFI technique, and the data protection that SFI provides is also needed in isolation applications.

\subsection{Branch Prediction}
In micro-architecture design, branch prediction is important to hide latency and allow instruction level parallelism and therefore improved performance. A poor branch prediction can harm performance as every  misprediction disrupts the  instruction pipeline.  Branch prediction includes predicting the direction and target of a branch. Direction prediction means anticipating whether the branch transfers control or falls through, while target prediction means determining where the control is going if the branch is taken. There are many prediction techniques discussed in the architecture literature, e.g \cite{ertlOpt, hennessyComputer, thangarajanSurvey}.

As  indirect branches are important in SFI, and we found it significant in some of our benchmarks (section \ref{eval}), we focus on indirect branch prediction. For direction prediction, indirect branches are statically predicted to be always taken. The most widely used target predictor for indirect branches is a \textit{branch target buffer (BTB)} \cite{ertlOpt}. A BTB is a cache-like structure indexed by  instruction  address. It stores the last used target address for a branch instruction. As the size of a BTB is limited, it can suffer from capacity and collision misses. In most modern CPUs (like Intel Sandy Bridge as our test machine) there is a   BTB for indirect branch target prediction (but the size and indexing function are not officially documented).

As return addresses are the most common form of indirect branches, a return-address stack \cite{webbSubroutine} is provided in CPUs as a separate structure to resolve return addresses early. Naturally it is a stack where on a function call the return address is pushed and on a function return the address is popped. The advantage of a return-address stack  over a BTB is evident  for functions that are called from multiple call sites. While the BTB just can predict the return from the  last call, a return-address stack can remember the returns as long as its size allows.

\section{Related Work}
\label{related}
\textbf{Software-Based Fault Isolation}
The notion of Software-Based Fault Isolation was first  introduced by Wahbe et al.~\cite{wahbeSFI}. Their idea of protecting code and data integrity was directing an unsafe instruction 
(i.e. indirect jump or memory write\footnote{This work and many subsequent ones consider data reads less harmful, so to avoid higher overhead
they tend to not cover secret data protection as a design goal.}) through a dedicated register and masking the register value appropriately 	 at  runtime. First they defined an unsafe
instruction as any instruction which jumps or writes to an address that cannot be resolved before execution begins. Then they divide  an application's address 
space into a code segment and a data segment where all addresses inside a segment have the same upper bits. The policy is that an untrusted application can  
execute  code only from its own code segment and  can write only into its own data segment. The implementation of such a policy using dedicated registers is straightforward: before any unsafe instruction, sandboxing  instructions are inserted to set the upper bits of the dedicated register to an appropriate  value and then the unsafe instruction
 is executed. As long as unsafe instructions are preceded by the sandboxing instructions, the execution is deemed safe. A verifier is provided to statically check that such an invariant is respected.

%

\textbf{PittSFIeld}
The original work on SFI was for RISC architectures, but on CISC architectures like x86, variable-length  instructions can  lead to many different instruction streams. An indirect jump can 
take control to any address in the code segment and if the address is not the start of an intended instruction, then a different stream of instructions  can be executed.
To tackle this challenge, PittSFIeld~\cite{mccamantPittS} imposes an artificial alignment on x86 instructions. Memory is divided into \textit{bundles} (called \textit{chunks} by the PittSFIeld authors) of size 16 bytes, and jump instructions 
are only permitted to target the beginning of a bundle. To enforce this property, every indirect jump instruction is preceded by a checking instruction to ensure the 4 lower
bits of the target address are zero. In addition, no instruction is allowed to cross bundle boundaries, and the pair of a checking instruction and the following jump, cannot
be split across two different bundles. McCamant and Morrisett proposed the use of no-ops to align the target of a jump at the beginning of a bundle. This way control flow cannot jump into the middle 
of an instruction and interpret a different and unintended stream of instructions. To ensure return address alignment, each \textit{call} instruction must be the last instruction 
of a bundle.



To avoid having the compiler and assembler in its trusted code base, PittSFIeld ensures binaries comply with the security policy via a static verifier. The verifier checks simple policies (e.g. no indirect branch instruction may be at the beginning of a bundle) before execution starts.

\textbf{Native Client}
Native Client~\cite{yeeNaCl} is a project developed by Google to let browser-based applications benefit from the computational speed of native code.  NaCl is a sandbox permitting a native x86 binary (x64 and ARM binaries in later versions)  to be executed as a browser plug-in. It provides operating system portability for  untrusted native code while  preserving system security  via data and code sandboxing. 

NaCl consists of two main parts: an \textit{inner sandbox} and an \textit{outer sandbox}. The inner sandbox is  similar to PittSFIeld in terms of constraining control flow using instruction alignment in 32-byte bundles. To reduce runtime overhead, the inner sandbox uses  x86 segmented memory to contain memory references. The outer sandbox is responsible for blocking  side effects by capturing system calls made by the process running the native module. System calls issued by untrusted native code are compared against a white list and only allowed ones are permitted.

As the inner sandbox of NaCl is  our interest in this project, we elaborate on it in more detail. The inner sandbox implements software-based fault isolation. Security rules are embedded  into a native binary via a modified version of the \textit{gcc} compilation toolchain. A static verifier validates that the rules are followed by the native code just before loading. As mentioned before, NaCl implements data sandboxing using the segmented memory mechanism provided by the x86-32 architecture. This way there is no need to put extra instructions to limit data load and stores. Using instruction alignment  at 32-byte bundles, NaCl makes sure that disassembling  the binary from start to end gives the only instruction  stream visible to the processor. Having such disassembly  allows unsafe instructions to be identified and prohibited. Some  such unsafe instructions include: \textit{syscall} or \textit{int} as the native code should not interact with OS directly, \textit{lds} as the native code is not allowed to update segment registers, and \textit{ret} as returning from function calls must be implemented by a sandboxed indirect jump. 

To constrain control flow, direct branches are computed statically and checked to target an instruction identified during disassembly. For indirect branches a combination of segmented addressing and address masking guarantees that the branch is sandboxed. Using the \textit{CS} register the address is constrained in the code segment and by masking the 5 lower bits of the branch target address it will be pointing to a bundle start. 

The operation of the validator is crucial for security. The NaCl validator is a small trusted library which is responsible for disassembling untrusted code and checking security policies. Starting from the first instruction, the validator first checks that it is a valid instruction, and that it is not crossing a bundle boundary. Then, if it is an indirect  branch, its preceding masking instruction must be in the same bundle. For the case of direct branches, the target address is calculated and stored in an array. If all of these checks pass, the instruction's address is stored as a valid address. At the end all addresses in the branch targets array are checked to be valid addresses.

The original proposal of NaCl was only for x86-32. In later research Native Client was extended to the x64 and ARM architectures~\cite{sehrNaclx64}. The most challenging part of these ports was the lack of the segmented memory feature of x86-32. Therefore the authors decided to abandon sandboxing load instructions and just use masked sandboxing  and guard pages on store instructions. Also in control flow containment, the high-order bits are masked, too. For the case of x64, dedicating a register to hold the  sandboxed address, and reserving large guard regions, helped in efficient implementation.

Because introducing bundles is a significant change to code layout, it
is most convenient to perform rewriting before assembly as part of the
compilation toolchain, when labels are still symbolic; NaCl uses a
modified version of the {\em gas} assembler.
Recent advances in tools that can recover symbolic labels from
stripped binaries~\cite{WangWW2015,Ramblr} could be used as a
pre-processing step to extend NaCl to off-the-shelf executables.

\textbf{RockSalt}
The original version of NaCl's verifier was a hand written program in C, responsible for ensuring sandboxing policies. As the policy enforcement is dependent on code disassembly, it is challenging to make sure the verifier is doing its job right. To address this issue Morrisett  et al.\@ proposed \textit{RockSalt}, \cite{morrisettRocksalt} a DFA-based verifier for NaCl. They built  a formal model of a large subset of the x86 instruction set and proved verifier correctness on it. Their DFA-style verification approach was also adopted into later versions of NaCl. 

RockSalt is notable for having a very detailed model of an SFI system, but some previous projects also applied formal methods to SFI. The PittSFIeld project~\cite{mccamantPittS} formalized 7 instructions of x86 and proved SFI policies are respected by those instructions.  Earlier Winwood and Chakravarty~\cite{winwoodProvably} provided a machine-checked proof using Isabelle/HOL for an instruction rewriting technique for RISC architecture. Their method replaces any indirect jump with a direct jump into a trusted dispatcher code.

\textbf{Monitor Integrity Protection}
In  more recent research, \textit{Monitor Integrity Protection} \cite{niuMIP} is proposed. Here the authors try to tackle the variable-length  instruction problem with a different approach. Instead of using no-ops to align instructions into fixed-sized chunks, they divide  the code region into variable-sized chunks and record the beginning  of each chunk in a bitmap named the \textit{chunk-table}. All branches are restricted to target only the beginning  of a chunk. In case of direct branches, such a property can be checked statically, and for indirect branches  a dynamic check looks up the chunk-table to ensure the branch target is valid. Compared to alignment-based techniques, \textit{MIP} is more space efficient (as no no-ops are used). Each \textit{MIP} object in addition to code and data region, contains a chunk-table. To support separate  compilation, when modules are combined,  their chunk-tables are merged. The authors implemented SFI using \textit{MIP} and showed that it has competitive performance with alignment-based SFI.

\section{\normalfont N\MakeLowercase{a}C\MakeLowercase{l} \scshape Allowing Cross-Bundle Instructions}
\label{cbi-nacl} 
Even though padding instructions (using no-ops) to aligned addresses is an effective solution to variable length instructions in CISC architectures, it has some drawbacks. The main disadvantage of inserting no-ops is harming instruction caching and prefetching. As an example, consider a loop in which all instructions in the loop can fit in the L1 instruction cache, but once padded, some portion of the loop does not fit in the L1 cache anymore. In this scenario, every iteration of the loop will cause instruction cache misses. Similarly instruction prefetch is less effective as the code space becomes sparse. As no-ops make actual instructions farther from each other, in each cycle prefetch loads fewer actual instructions from the memory. 

Another drawback of no-op instructions is wasting CPU cycles once such instructions are loaded and executed. As an optimization, NaCl will jump over any sequence of 16 bytes or longer of no-ops, to avoid CPU cycle waste. In addition, padding the binary increases its size; this is not desirable  especially  for NaCl where native modules need to be first downloaded on a client machine and then  executed. An increase in binary size will result in an increase in download time (though less with compression). 

Considering these drawbacks encouraged  us to look for possible ways to reduce the number of no-ops used for padding. So as the main idea of this project, we try to remove unnecessary no-ops while still satisfying the same safety requirements as in vanilla NaCl.

\begin{algorithm}[!t]

\begin{algorithmic}[1]
\Procedure{pad\_removal}{\textit{assemblyFile}, \textit{padInfo}}
\ForAll{$pad$ \textbf{in} \textit{padInfo}} 
\State $val \gets \text{FALSE}$
\State $savedSize \gets pad.size$
\State $succPad \gets \textit{padInfo}.successor(pad)$
\State $savedSuccSize \gets succPad.size$
\State $pad.size \gets 0$
\While{$pad.size \leq savedSize$}	
\If{$succPad$}
\State $\textit{diff} \gets savedSize~-~pad.size$
\State $succPad.size \gets$ 
\\ \hfill $(savedSuccSize + \textit{diff}) \mathbin{\%} 32$
\EndIf
\State $nexe \gets \textsc{build}(\textit{assemblyFile}, \textit{padInfo})$
\State $val \gets \textbf{validate}(nexe)$
\If{$val = \text{TRUE}$}
\State \textbf{break}
\EndIf
\State $pad.size \mathrel{+}= 1 $
\EndWhile
\EndFor
\EndProcedure 
\Procedure{build}{\textit{assemblyFile}, \textit{padInfo}}
\State $objFile \gets \textbf{assemble}(\textit{assemblyFile}, \textit{padInfo})$
\State $nexe \gets \textbf{link}(objFile)$
\State \textbf{return} $nexe$
\EndProcedure
\end{algorithmic}
\caption{Greedy Pad Removal}\label{pad_removal}
\end{algorithm}

\subsection{Changing NaCl Padding Scheme}
\label{sec:pad_rm}
In vanilla  NaCl, an instruction will be padded in three cases: 1) if it is an indirect jump target, 2) if it is a call, or 3) if it  crosses a bundle boundary. In case 1, the instruction is aligned at the beginning  of the next 32-byte bundle. In case 2, the call instruction is padded to be positioned at the end of a bundle. In  case 3, the instruction is padded to the beginning  of the next bundle. Cases 1 and 2 are necessary for sandboxing to preserve the code's behavior. Since the target of an indirect jump is masked so that it will point to a 32-byte-aligned address,  the target instructions must be positioned at bundle starts. In case 2, the instruction after the call is the target of return instruction, so pushing the call to the end of the bundle positions the target instruction at the next bundle start. Case 3 pads instructions which would otherwise start in one bundle and end in the next bundle. This type of padding is to eliminate   unintended or invalid instructions. But it is conservative; we could allow a cross-bundle instruction if we make sure  no unsafe instruction is interpretable from the crossing point. In this way we can decrease the number of no-ops. From now on we use \textit{Cross-Bundle Instruction (CBI) NaCl} to refer to our changed version of NaCl which relaxes the requirements of case 3.

The main challenge is deciding which no-ops can be  removed while ensuring the binary still passes the verifier. It is an optimization problem, which for now we use a greedy method to solve. For simplicity our approach reuses the validator with its standard interface. (As future work we would like to explore other algorithms, mentioned in Section~\ref{future}.) In our method, for each padding of type 3 (cross-bundle instruction), we start by removing the whole padding sequence, and iteratively we increase the padding size by one byte and check the effect on the validation. If validation succeeds, it means no-op removal did not create any invalid instructions, so we can replace the current padding with a shorter one, and go to the next padding. Algorithm \ref{pad_removal} shows the pseudocode for greedy pad removal.

For each source code file, we retrieve assembly code using the \textit{-S} switch to \textit{gcc}. We also made a change in the NaCl build chain to extract information related to pad placement. The modified assembler produces a sequence of entries each of which  consists of a  location (saying which instruction is  padded), and a size (the length of the pad in bytes) which are collected into a data structure named \textit{padInfo}.

In Algorithm \ref{pad_removal} the loop at line 2 keeps this invariant that at the start of each iteration the \textit{padInfo} leads to a valid binary. 
For each element in \textit{padInfo} we start by setting the size of padding to zero and increase it one by one if necessary. In order to evaluate just the effect of current padding size change, the size difference will be added to the next element in the \textit{padInfo} (line 10). 
This way, in the inner loop (line 8) the whole binary will be like the previous step except for the current padding size change. 
Note that the addition is modulo bundle size as we can remove any full bundle of no-ops.
Another perspective is that this approach moves padding instructions towards the end of the file instead of removing them directly.
But in fact, in each iteration the accumulated pads are modulo 32, which means any pads over 32 will be shrunk, 
and also at the end of the file, any remaining pads can be safely removed without changing the binary's validity.


\subsection{Multipass Validator}
The NaCl validator checks instructions in one pass starting from the beginning of the code and going to  the end. The validator maintains two bitmaps: one for safe addresses to jump to (named \textit{valid targets}), and one for actual addresses which the code will jump to (named \textit{jump targets}). For each instruction, it checks whether it is a valid instruction or not. If it is valid, then  the instruction address is marked in the \textit{valid targets} bitmap. If the instruction is a direct branch, the target address is marked in the \textit{jump targets} bitmap. In the case of an indirect branch, the validator checks that  it is preceded by appropriate masking instructions. At the end of the validation pass, the two bitmaps \textit{valid targets} and \textit{jump targets} are compared  to make sure all direct jumps are targeting valid instructions. 

Considering our changes in the NaCl padding scheme, we need to make changes in the validator to cover all possible instruction streams. As we are allowing cross bundle instructions, we must make sure that any instruction stream starting from the crossing point will not reach an invalid instruction. To do this we validate a binary in multiple passes. Each validation pass starts from a bundle start (as the bundle start is the possible crossing point); and in each pass, the validation is the same as vanilla NaCl. Algorithm \ref{multipass} presents the multipass validator pseudocode which is an adaptation of RockSalt's validator \cite{morrisettRocksalt}.
The changes from the RockSalt algorithm are highlighted in gray.

\begin{algorithm}[!t]
\centering
\algdef{SE}[CWHILE]{CWhile}{CEndWhile}[1]{\colorbox{gray!20}{\algorithmicwhile\ #1\ \algorithmicdo}}{\colorbox{gray!20}{\algorithmicend\ \algorithmicwhile}}%
\algtext*{CEndWhile}
\begin{algorithmic}[1]
\Procedure{validate}{$code$, $size$}
\State initialize two arrays of size $size$ named $valid$ and $target$ to FALSE
\State $bundleStart \gets 0$
\State $result \gets \text{TRUE}$
\CWhile{$bundleStart < size$}
\State $pos \gets bundleStart$
\While{$pos < size$ \colorbox{gray!20}{\&\& $!valid[pos]$}}
\State $valid[pos] \gets \text{TRUE}$
\State $savedPos \gets pos$
\If{\textbf{match\_maskedJump\_insn}( $code$, $\&pos$, \\ \hfill  $size$)}
\State \textbf{continue}
\EndIf
\If{\textbf{match\_nonControlFlow\_insn}($code$,\\ \hfill $\&pos$,  $size$)}
\State \textbf{continue}
\EndIf
\If{\textbf{match\_directJump\_insn}($code$, $\&pos$, \\ \hfill $size$)}
\State $dst \gets$ \textbf{extract\_dest}($code$, $savedPos$,\\ \hfill $pos$, $size$)
\State $target[dst] \gets \text{TRUE}$
\State \textbf{continue}
\EndIf
\State \textbf{return} FALSE
\EndWhile
\State \colorbox{gray!20}{$bundleStart$ += $32$}
\CEndWhile
\For{ $i=0$ to $size$}
\If {$target[i]$ \&\& $!valid[i]$}
\State $result \gets \text{FALSE}$
\EndIf
\If{ $i$ is 5-bit aligned $\&\&$ $!valid[i]$}
\State $result \gets \text{FALSE}$
\EndIf
\EndFor
\State \textbf{return} $result$
\EndProcedure
\end{algorithmic}
\caption{Multipass Validator}\label{multipass}
\end{algorithm}

The Validate procedure takes two parameters, a pointer to the code location, and the size of the code. At line 2, two arrays (\textit{valid} and \textit{target}) of size of the code  are initialized to FALSE. The \textit{valid} array is used to mark all valid instructions addresses (which in turn can be valid jump destinations). The \textit{target} array records those addresses that are jumped to by some direct jump or call instructions. 

Two nested while loops at lines 5 and 7 are responsible for going through the code in multiple passes and  validating  all possible addresses which can be the start of an instruction. The while loop at line 5 iterates over all aligned addresses (bundle starts), and in each iteration it validates instruction streams beginning from the bundle start. The inner loop at line 7, executes  if the address in \textit{pos}  has not already been validated. If it was, it means the instruction stream was seen in previous iterations and there is no need to be re-validated. Otherwise,  for each address in \textit{pos}, the algorithm tries to match the  bytes in \textit{code} starting from \textit{pos}  against the three classes of instructions (masked jump, non-control flow, and direct jump). Once matched, \textit{pos} is advanced by the size of matched instruction and the process goes to the next iteration. 

The \textit{match\_maskedJump\_insn} procedure matches  a masking instruction on register \textit{r} followed by an indirect call or jump through \textit{r} (line 10). The \textit{match\_nonControlFlow\_insn} procedure matches any instruction allowed by NaCl which does not affect control flow (line 13). The \textit{match\_directJump\allowbreak\_insn} procedure matches direct call or jump instructions (line 16). When this procedure is matched, the \textit{extract\_dst} procedure extracts the target of the direct jump or call (line 18), and the destination is recorded in the \textit{target} array (line 20). When the while loops finishes, the algorithm checks at line 25 that each target address is a valid address, and at line 27, it checks that each aligned address is the beginning of a valid instruction.

The vanilla NaCl validator processes a binary in linear time
(proportional to the code size).
Making multiple passes makes validation more expensive, but it is
still asymptotically linear.
It might seem the nested loops would have quadratic complexity.
However there can be a limited number of distinct execution streams in
the same code bytes: eventually most streams will have to converge.
Once an execution stream converges with an already-validated
instruction stream, the rest of the already-validated stream does not
need to be rechecked.
Algorithm \ref{multipass} implements this optimization by checking the
flags in the {\em valid} array.
If the {\em valid} flag for the current address is already set, the
validation of the current stream stops and we return to the outer
loop.
Each iteration of the inner loop marks a previously-unseen
instruction, and instructions are never unmarked, so every byte of the
code will be processed (as start of an instruction) at most once.
The worst-case number of instruction checks increases from the number
of intended instructions in the code, to the number of byte locations
in the code, but the validator running time is still linear in the
binary size.

The optimization of not rechecking already-validated instructions does
not change the validator result because all of the operations
performed when checking an instruction are idempotent.
In other words, performing them once has the same effect as performing
them repeatedly: the matching result depends only on the instruction
bytes, which do not change, and once flags in the {\em target} array
are set, they stay set.
The sequence of remaining instructions in a stream is uniquely
determined from any address in the stream.
When the algorithm encounters an already-processed instruction, it is
equivalent to skip the checking of the rest of the stream, because all
of the result of that checking have already been recorded.
The \textit{match\_maskedJump\_insn} check treats two x86 instructions
as one pseudo-instruction, where jumping directly to the second of the
two instructions would be unsafe.
The algorithm treats this in the same way as an unsafe instruction
found by interpreting bytes within a single x86 instruction: the
second instruction is not valid on its own, so validation will fail if
it is reachable in any other way, including by a direct jump, by
occurring at a bundle boundary, or along an overlapping-instruction
stream.

\subsection{Correctness of the Multipass Validator}
After describing the changes made in NaCl, here we argue why our changes do not violate NaCl sandboxing  policies. The NaCl validator is responsible for  policy enforcement. If we can show that despite our changes, the validator still enforces the sandboxing policies correctly, then we are done.

The RockSalt~\cite{morrisettRocksalt} authors proved the correctness of their validator which in turn reflected in the current NaCl validator. So we can assume that the NaCl validator is correct and will argue that our changes on making the validation process multipass does not disrupt the validator correctness. The proof in \cite{morrisettRocksalt} relies on the fact that during the binary execution contents of segment registers are not changed from the initialization moment and also the bytes  in the code segment are not altered after being validated. 

The validator prevents untrusted code from updating segment registers by filtering such instructions. As we did not change set of forbidden instructions, therefore the multipass validator provides same protection. This way, if the untrusted binary is loaded in an appropriate environment as long as the instruction pointed to by \textit{pc} (program counter register) is valid, the sandboxing policy is satisfied. The multipass validator covers all addresses that can ever be an instruction address. That is because for the case of direct jumps, the loop at line 20 of Algorithm \ref{multipass} makes sure jump destinations are valid instructions; and for the case of indirect jumps (as they are just allowed to target aligned addresses) the outer loop at line 5 covers all aligned bundle starts. 
So based on our changes, if the multipass validator returns \textit{TRUE} on a binary, it means in every pass staring from a bundle start, the validator deemed the instruction stream to be in accordance with the sandboxing policy. 

To get a machine-checked verification of our reasoning, we have also
formalized the above argument in the Coq~\cite{Coq} proof assistant as
a modification to the previous RockSalt~\cite{morrisettRocksalt}
proof.
The original RockSalt proof modeled a single-pass validator, so we
modified it to model a multipass validator as in
Algorithm~\ref{multipass}.
Loops are modeled as recursive functions in Coq, and we
reused the function implementing the loop of the single-pass validator
as the inner loop of the multipass validator (the model does not
include the optimization of exiting the inner loop early when
encountering an already-validated instruction).
We then inserted a new recursive function corresponding to the outer
loop that starts a verification pass at each bundle start address.
To allow the rest of the proof to work with this change, we had to
inductively prove several properties about the outer loop that were
already proved for the inner loop, for instance that when an
instruction was added to the {\em valid} bitmap, it was within the
range of the code region and was in fact the address of a valid
instruction.
In total we changed about 20 lines of the model and wrote about 650
lines of new lemmas and their proofs.
%
The proof is available at \url{https://pastebin.com/gN026Hte}.
With these changes, the security guarantee proved for RockSalt applies
with the same force for our modified approach, so we have a strong
assurance that allowing cross-bundle instructions does not decrease
security.

\section{Evaluation}
\label{eval}
To evaluate our changes to Native Client we used the SPEC CPU suite (a standard set of CPU-intensive programs). We began with SPECint2000 as it was used in the NaCl papers \cite{yeeNaCl, sehrNaclx64}. We then decided to move on to SPECint2006 as it consists of benchmarks with bigger code sizes. We expected our CBI NaCl to show better performance on bigger binaries that have higher instruction-cache pressure.
(We also looked into the recently-released SPEC CPU 2017 suite, but it appears to have more severe porting obstacles related to Native Client's older GCC version and memory usage limits.)
We used GCC version 4.4.3, and the test machine was an Intel Core i7-3770 (Sandy Bridge) CPU at 3.40GHz, with 16GB of memory, running Ubuntu 16.04. 

We made a virtual machine containing the CBI NaCl software (source code, compiled binaries, and scripts) which can be accessed at 
\url{https://www-users.cs.umn.edu/~emamd001/cbi-nacl.html}.

\begin{figure*}
\centering
\includegraphics[width=0.95\textwidth]{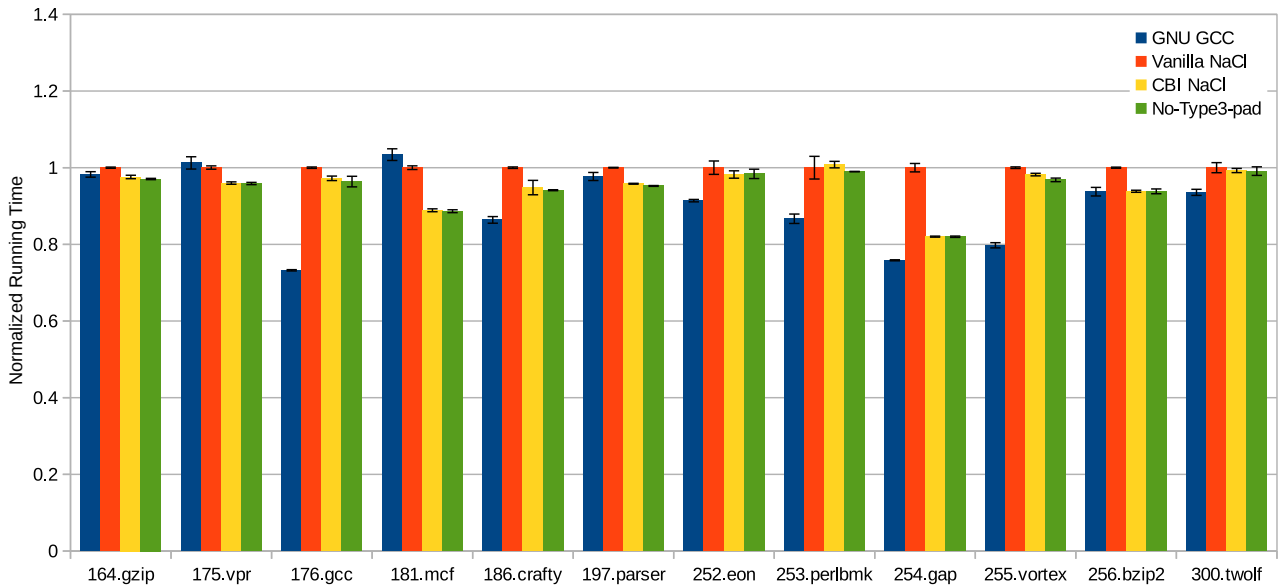}
\caption{Running time overhead comparison for SPECint2000 (normalized with respect to vanilla NaCl)}
\label{fig:runtime}
\end{figure*}

\subsection{Results for SPECint2000}
\label{sec:result2000}
The SPECint2000 package consists of 12 programs (11 written in C and one in C++). We built these benchmarks on x86-32 architecture and measured the running time of each sample. 
Figure \ref{fig:runtime}  shows the running time overhead comparison for the SPECint benchmarks as built by \textit{GNU GCC}, vanilla NaCl, and CBI NaCl compilers. Each benchmark was run 11 times, and after discarding the first  result, the average running time for 10 runs was reported.
The best case overhead reduction is for \textit{gap} with 17.9\% and on average CBI NaCl binaries show 4.78\% improvement in running time.
For the case of \textit{perlbmk} CBI NaCl binary we observed an increased runtime overhead of 0.8\%. This increased overhead is unexpected, and we describe a more detailed investigation in \ref{anomaly}.


Table \ref{tab:filesize} shows the code size of the benchmarks for each compiler in bytes. As it can be seen, except in two cases, our changed compiler generated a smaller binary compared to vanilla NaCl. This size difference is due to pad removal, as we did not change the  data layout of binaries and only changed the code layout. For the cases of \textit{181.mcf} and \textit{252.eon} we see a small increase in the binary size of CBI NaCl. That is because at the time of compilation, we put a bundle (32 bytes) of NOP instructions at the end of each object file. These extra bundles are placed to merge the validation paths once all the object files are linked together to build the final binary. For these two benchmarks, the size of the extra bundles exceeds the number of bytes removed from the padding. But as these NOP bundles are placed after the last instruction of the last function of each object file, they will not be executed and have no effect on the number of actually executed instructions.

\begin{table}
\centering
\caption{Code size for SPECint CPU2000 in bytes.}
\label{tab:filesize}
\begin{tabular}{|c||r|r|r|} \hline

\textbf{Benchmark}&\textbf{GNU GCC}&\textbf{Vanilla NaCl}&\textbf{CBI NaCl} \\ \hline \hline
   
164.gzip&97,816&244,205&243,889\\ \hline  
175.vpr&254,574&622,047&621,683\\ \hline  
176.gcc&2,030,316&5,598,520&5,598,512\\ \hline 
181.mcf&29,405&92,248&92,464\\ \hline
186.crafty&333,524&699,416&699,176\\ \hline  
197.parser&276,651&784,118&782,806\\ \hline  
252.eon&575,211&4,840,324&4,844,416\\ \hline 
253.perlbmk&829,860&2,444,380&2,441,480\\ \hline 
254.gap&956,543&2,273,206&2,268,190\\ \hline 
255.vortex&754,974&2,070,729&2,067,929\\ \hline 
256.bzip2&96,640&207,686&206,438\\ \hline  
300.twolf&415,700&1,056,335&1,055,975\\ \hline
\end{tabular}
\end{table}

Table \ref{tab:insn} shows the number of instructions executed by each benchmark at  runtime. These numbers are collected by running each sandboxed benchmark under Valgrind/Cachegrind tool. As  shown, our pad removal decreased the number of instructions executed. The removed instructions are the padding instructions which have no effect on the  benchmark's functionality.

\begin{table}
\centering
\caption{Number of runtime instructions for SPECint CPU2000.}
\label{tab:insn}
\begin{tabular}{|c||r|r|c|} \hline
&\textbf{Vanilla NaCl}&\textbf{CBI NaCl}& \!\small{Decrease}\!\! \\ \hline  \hline    
164.gzip&438,584,902,313&429,015,758,807&2.8\%\\ \hline  
175.vpr&200,396,220,469&200,384,630,579&\textless0.1\%\\ \hline	
176.gcc&175,502,236,335&174,171,040,934&0.7\%\\ \hline 
181.mcf&55,730,276,258&54,432,114,188&2.3\%\\ \hline
186.crafty&226,135,233,139&223,841,297,154&1.0\%\\ \hline  
197.parser&342,435,548,778&340,426,993,766&0.5\%\\ \hline 
252.eon&215,902,254,049&215,902,222,993&1.4\%\\ \hline	
253.perlbmk&400,334,639,815&396,991,025,588&0.8\%\\ \hline 
254.gap&248,434,693,423&245,972,158,831&1.0\%\\ \hline 
255.vortex&380,247,713,301&374,391,454,899&1.5\%\\ \hline 
256.bzip2&348,260,041,064&343,301,990,583&1.4\%\\ \hline  
300.twolf&351,235,815,588&347,811,831,402&1.0\%\\ \hline
\end{tabular}
\end{table}

We should mention that as we are using a greedy method for pad removal, the result may be sub-optimal and therefore the overhead reduction may not be the best possible. But still these results confirm the effect of pad removal on improving performance.
To determine an upper bound on how much performance benefit we can get by any approach to remove type 3 padding (such as replacing our current greedy algorithm with a globally-optimal one), we did the following experiment. We recompiled all 12 SPECint2000 benchmarks with the NaCl build chain but removed all type 3 padding. This is not secure and the generated binaries are rightfully rejected by the NaCl validator, but we can still measure their performance overhead. We executed these generated binaries 11 times and reported the average running time of the last 10 executions. The fourth column in figure \ref{fig:runtime} depicts this result. This experiment showed that type 3 pad removal yields about 5.3\% overhead reduction on average. Comparing this result to CBI NaCl running time confirms that our current approach achieves more than 90\% of the possible running time improvement.

\subsection{Results for SPECint2006}
As the performance improvements for SPECint2000 were unsatisfying, we decided to next consider the newer SPECint2006 suite. We believed as most of the SPECint2000 benchmarks have relatively smaller code sizes, they could fit mostly in current CPU instruction caches. So we were interested to see how using a set of benchmarks with bigger code sizes could affect the performance improvement. 

SPECint2006 consists of 12 benchmarks (9 in C, and 3 in C++). We managed to build 10 of these benchmarks and measured the running time for 4 different compile setups. We could not successfully execute one of these benchmarks (\textit{400.perlbench}) inside the sandbox (neither vanilla NaCl, nor CBI NaCl) and one other benchmark (\textit{483.xalancbmk}) failed even to compile using g++ 4.4.3. 

\begin{figure*}[!t]
\centering
\includegraphics[width=.95\textwidth]{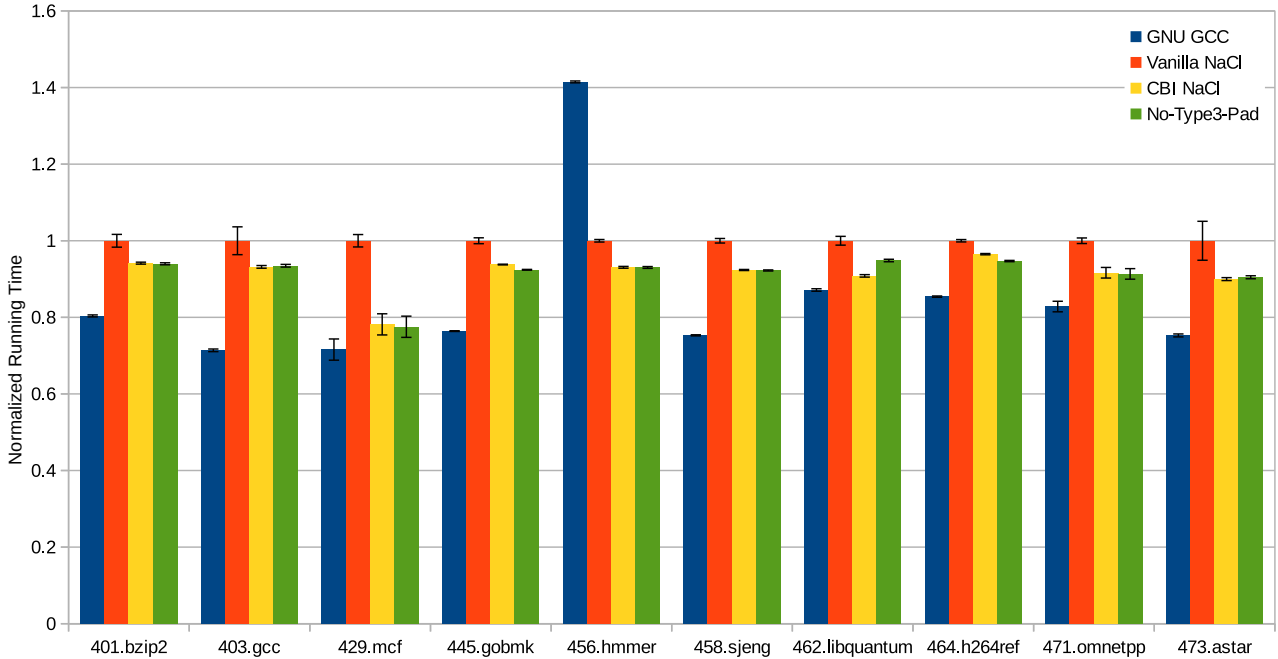}
\caption{Running time overhead comparison for compatible benchmarks
from SPECint2006 (normalized with respect to vanilla NaCl)}
\label{fig:runtime2006}
\end{figure*}

Figure \ref{fig:runtime2006} shows the running time overhead comparison of SPECint2006 benchmarks. Each benchmark was run 11 times, and
after discarding the first run, the average running time of 10 runs was reported.
The average overhead reduction is 8.6\% which shows a 
meaningful performance improvement compared to \ref{sec:result2000}. As we mentioned before, we hypothesize that the bigger code sizes of SPECint2006 
are a significant reason for such differences. The best case overhead reduction is for \textit{mcf} with 21.8\%. 

Table \ref{tab:insn2006} depicts the number of instructions executed  of benchmarks for each compiler. For this experiment we were not able to monitor SPEC2006 benchmarks under Valgrind because these benchmarks required a larger address space to run, conflicting with Valgrind's memory allocation scheme. Therefore we used OProfile (a statistical profiler for Linux binaries)~\cite{OProfile} to estimate the number of retired instructions with a sampling rate of $10^6$. This table shows the CBI binaries executed on average 15\% fewer instructions at runtime.

\begin{table}
\centering
\caption{Number of runtime instructions for SPECint CPU2006.}
\label{tab:insn2006}
\begin{tabular}{|c||r|r|c|} \hline
&\textbf{Vanilla NaCl}&\textbf{CBI NaCl}& \!\small{Decrease}\!\! \\ \hline  \hline    
401.bzip2&3,379,347 $\times 10^6$&2,917,477$\times 10^6$&13.6\%\\ \hline
403.gcc&1,799,683$\times 10^6$&1,402,218$\times 10^6$&22.0\%\\ \hline
429.mcf&563,035$\times 10^6$&423,872$\times 10^6$&24.7\%\\ \hline
445.gobmk&2,400,400$\times 10^6$&1,957,218$\times 10^6$&18.4\%\\ \hline
456.hmmer&4,452,957$\times 10^6$&3,564,759$\times 10^6$&19.9\%\\ \hline
458.sjeng&3,276,737$\times 10^6$&2,734,397$\times 10^6$&16.5\%\\ \hline
462.libquantum&2,992,638$\times 10^6$&2,974,543$\times 10^6$&0.6\%\\ \hline
464.h264ref&4,688,793$\times 10^6$&3,991,415$\times 10^6$&14.8\%\\ \hline
471.omnetpp&943,026$\times 10^6$&846,961$\times 10^6$&10.1\%\\ \hline
473.astar&1,625,798$\times 10^6$&1,491,833$\times 10^6$&8.2\%\\ \hline

\end{tabular}
\end{table}

\subsection{Investigating the Anomaly}\label{anomaly}
As figures \ref{fig:runtime} and \ref{fig:runtime2006} show, in some cases, pad removal unexpectedly caused a slightly increased overhead. To investigate the cause, we investigated more detailed profiling of the benchmark execution.
These results suggest address-dependent indirect branch prediction as
a source of performance variation.

We first made some changes in the NaCl (described in section \ref{challenge}) to be able to run under Valgrind/Cachegrind. 
Though Cachegrind has the advantage of not requiring statistical
sampling, its results did not seem very predictive of performance
overhead.
From further investigation we hypothesize that this is because its
simulation of indirect branch prediction is currently relatively
simplistic: it uses a direct-mapped cache indexed by the low 9 bits of
an instruction address, which is much simpler than the BTB in a modern
CPU.

Therefore we used OProfile  to estimate some interesting events for each benchmark execution (events include the number of instructions retired, instruction cache misses, branch misprediction,  etc). From the OProfile logs we observed there is an increase in the number of branch mispredictions for the case of our CBI NaCl compiled binaries. So we hypothesized that our pad removal had an unpredictable side effect on indirect branch prediction success. That could be the case because once we remove padding, instruction addresses will change. Indirect branch prediction (via the BTB) is dependent on the address of indirect branch, so changes to collisions in the BTB  may increase the miss rate.

A small change in a program can perturb its layout which affects the cache and branch predictors in complex ways.
To confirm the effect of branch misprediction on runtime, we took an approach like one proposed by Curtsinger et al.~\cite{curtsingerStabilizer} to generate multiple samples of the SPECint2000 \textit{perlbmk} benchmark with randomized layout.
 We changed the way the assembler places indirect call instructions in a bundle. One layout is like vanilla NaCl, meaning that an indirect call instruction is placed at the end of current bundle. Another layout placed the  indirect call instruction at the end of a new bundle. Then, the assembler randomly selects one of these layouts for each indirect call instruction. This way we can produce different versions of same benchmark binary with different instruction address layout. 

\begin{figure*}[!t]
\centering
\subfloat[Fitting]{\includegraphics[width=0.8\textwidth]{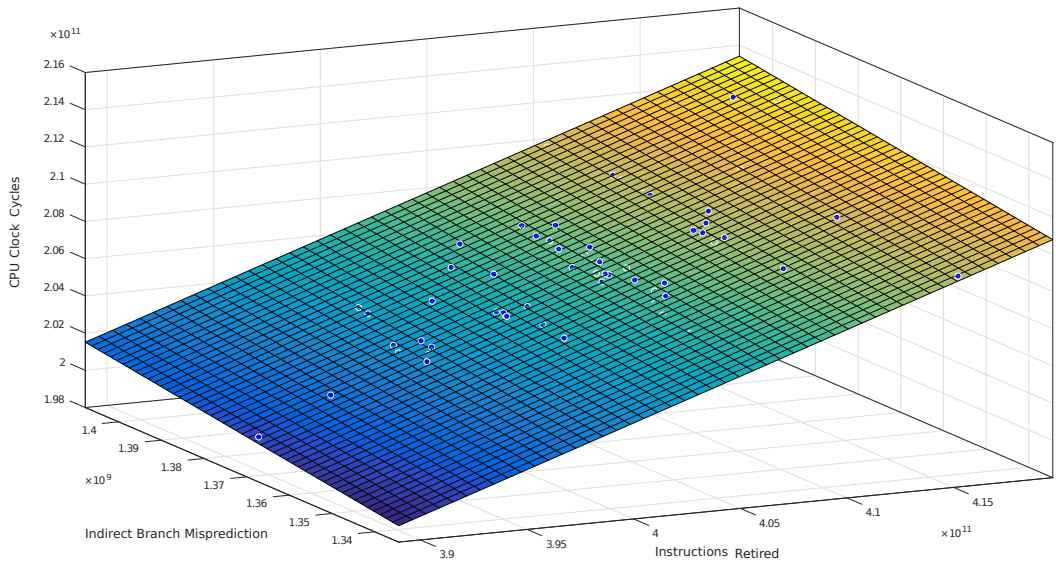}
\label{a}}
\hfil
\subfloat[Residual]{\includegraphics[width=0.8\textwidth]{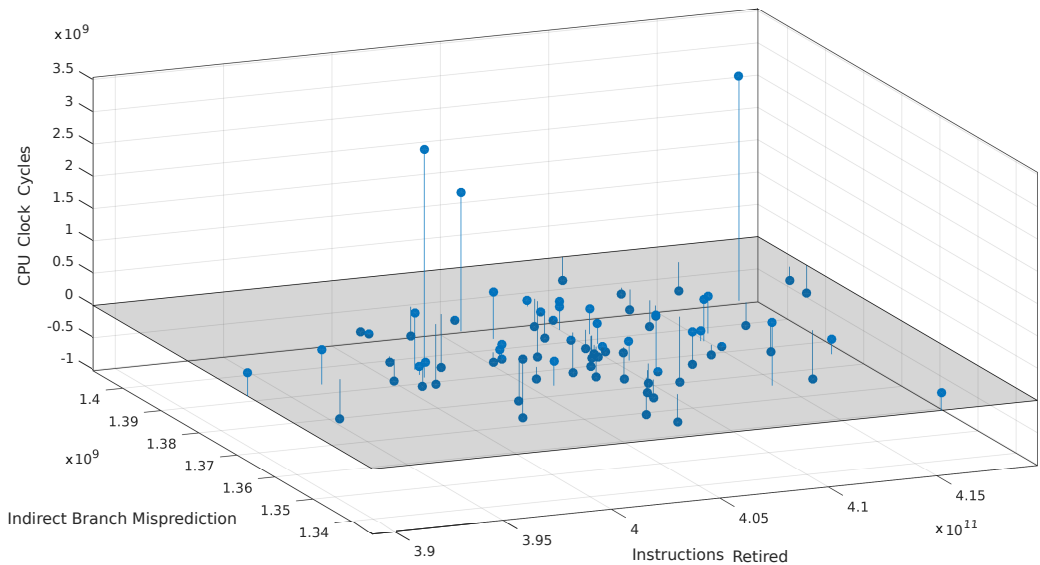}
\label{b}}
\caption{Correlation between CPU Clock Cycles and Indirect Branch Misprediction for  \textit{perlbmk}}
\label{matlab-fit}
\end{figure*}

We build 80 different versions of \textit{perlbmk} binaries (for vanilla NaCl and CBI). Then we monitored these samples' execution under OProfile and collected the number of instructions retired, number of indirect branch mispredictions, and CPU clock cycles.
To separate the effects of changing the number of instructions
executed and the number of mispredicted branches, we fit a linear
relationship shown in (\ref{formula}) between these variables.
The coefficient $x$ is the number of cycles per instruction, while $y$
is the penalty in cycles for a mispredicted indirect branch.
\begin{small}
\begin{equation}
\label{formula}
(\text{CPU cycles}) = x\times(\text{\# of insns}) + y\times(\text{\# of mispredictions})
\end{equation}
\end{small}
These variables turn out to be strongly related; the linear fit is
shown graphically in Figure~\ref{matlab-fit}.
With these three variables the $R^2$ statistic is 0.904; $x$ is 0.388
CPI, while $y$ is a 35.8 cycle misprediction penalty.
By comparison, if the mispredictions are omitted, the $R^2$ is only
0.796.

These results suggest that the cause of the anomaly was an increased number of indirect branch mispredictions. In some cases, pad removal may affect the effectiveness of the BTB by changing the instruction address layout. To avoid this kind of disruption, one might incorporate instruction re-ordering to place indirect branch instructions at different addresses. We leave investigating such remedies for future work.

\subsection{What about x64?}

The CBI NaCl approach is also applicable to the x64 architecture;
unfortunately it provides much less benefit there.
The main reason is that the x64 validation rules allow fewer
instructions.
Under x86-32 NaCl, most instructions that access memory are allowed
because the address will be checked with segmentation at runtime.
Under x64 NaCl, by contrast, only a restricted set of addressing modes
are legal, and because of this, fewer overlapping instructions can be
allowed.
To isolate and measure this effect in more detail, we performed an
experiment with random byte sequences.
(Though overlapping instructions are not truly random, they are
outside our tool's control.)
We generated 1 million bundle-length byte sequences, and checked
whether each sequence was legal under the rules of the x86-32 or x64
validators.
On average 9.1\% of random byte sequences are allowed on x86-32, while
only 0.37\% are allowed on x64.

We have also implemented CBI NaCl for x64 and evaluated its
performance effect.
However the performance benefit is small, averaging less than 0.5\%.
Examining the binaries, our system was rarely able to remove padding,
confirming the effect suggested by the random-byte-sequence experiment.

\section{Challenges}
\label{challenge}
In this section we talk about further implementation details of the project which were necessary to accomplish our main goals.

\subsection{On Generating Valid NaCl Binaries}
\label{sec:gen-nexe}
As mentioned in Section \ref{sec:pad_rm}, Algorithm \ref{pad_removal} processes each source code file one by one and tries 
to remove padding as much as possible, while maintaining the invariant that the object file passes validation.
Then these object files are linked into a complete \textit{nexe} executable.
Maintaining this standard C separate compilation approach in padding removal is helpful for scalability, but it requires that the linking step not break validation.

The assembler is the tool that puts instructions in the object file and determines the relative address of instructions. 
We implemented the padding removal by changing the way \textit{gas} puts padding into the object file. The binary content is the 
same from object file to final \textit{nexe} except for the relocation entries which are filled in by the linker. Once pad removal 
decides appropriate padding for all source code files, their corresponding object files are linked together to generate the final 
\textit{nexe}, and this final \textit{nexe} is passed to the validator to make sure the generated \textit{nexe} is a valid one. 
The final linking step is where relocations (like function names or jump destinations) are resolved to actual  addresses inside 
the binary. This means relocations are replaced by addresses, so the byte stream which the final validation sees is different from 
the streams validated in intermediate validations. So if we are not careful, it may happen when the relocations are replaced by 
actual values, the changed bytes may be interpreted to form an unsafe instruction. 

To avoid such situations, we wish to be conservative toward labels and avoid having paths with labels crossing the bundle boundaries. 
Conceptually, we would like the validation of a single object file to fail if any value for a relocation address would cause validation to fail.
To approximate this check using the existing validator, we replace labels with byte values that are most likely to cause validation to fail if they are interpreted as an instruction.
Specifically, in intermediate steps we replace labels with the hex bytes \texttt{c3 c3 c3 c3}.
\texttt{0xc3} is the opcode for a one-byte \texttt{ret}
instruction, which is prohibited by the validator.
Thus if any location inside the label could be reached as an
instruction, we conservatively reject the layout.
(The change is made only for the purposes of validation testing, not
in the object file used for linking, because the label bytes can
contain information used by the linker.)

However this approach is not quite complete: for instance in SPECint2000, it fails for two benchmarks (\textit{crafty} and \textit{vortex}).
The failure occurs when a byte from a label is interpreted not as an opcode but as a mod-r/m byte, which affects the length of an instruction and in turn the stream of later instructions.
We handled this case by generalizing the \texttt{0xc3} approach to allow re-testing per-object validation with other byte values as well.
Since this problem is relatively rare, our system first attempts to compile a binary testing with just \texttt{0xc3}.
If validation fails after linking, our system extracts the label byte value that led to the validation failure, adds it to the testing set, and retries the compilation.
For SPECint2000 \textit{crafty} and \textit{vortex}, final validation succeeds with the addition of \texttt{0x9f} or \texttt{0xbf}   respectively.
While building SPECint2006, the tool found other filtered bytes for three more benchmarks. It found 
 \texttt{0xa0}  for \textit{gcc}, \texttt{0xe0} for \textit{gobmk},  and both  \texttt{0xa0} and \texttt{0x20} for h264ref. 

\subsection{On Adapting Valgrind  to Run on NaCl}
As we mentioned in section \ref{eval}, it was appealing to us to be able to monitor a NaCl sandboxed binary under Valgrind: the Valgrind-based \textit{cachegrind} tool simulates a binary's interaction with the system cache and branch predictor), and Valgrind is also useful for other kinds of debugging.
The NaCl project provides scripts to let developers  run Valgrind/Memcheck and ThreadSanitizer on x64 versions of NaCl binaries. 
But our need was to run Valgrind's Cachegrind on x86-32 binaries.

The way NaCl loads the sandboxed binary into the memory, and the use of segment-based addressing are the main obstacles to 
running Valgrind/Cachegrind on NaCl. NaCl first allocates a memory region with a random base address, then initializes the 
segment registers appropriately, and then uses an \texttt{ljmp} instruction to jump into the sandbox. 

We decided to disable segment-based addressing in  NaCl during Valgrind debugging. At a high level, the changes we made are 
that first we fixed the base address that NaCl loads the binary into, and then report address zero as the  virtual address. 
In this way  the base address of the segments will become zero (i.e. there is no need to use the base address of segment registers  
for correct addressing). We then located the NaCl code which updates segment registers and transfers the control into and 
out of the sandbox. We replace this code with equivalent instruction that do not change segments. For example we replace 
every usage of the \texttt{lss} instruction (which loads \texttt{\%esp} and the stack segment register \texttt{\%ss}) with 
regular \texttt{mov} instruction to load an offset into  \texttt{\%esp}. As another example we replaced every instruction 
working with \texttt{\%cs} (like \texttt{ljmp} or \texttt{lcall}) with a regular \texttt{jmp} or \texttt{call} via the offset.

With these changes we were able to load and execute a sandboxed binary with  NaCl under Valgrind. Using the Cachegrind tool, 
we were able to count the number of instructions executed by each benchmark at  runtime (Table \ref{tab:insn}).

Even though with the aforementioned changes we were able to execute NaCl under Valgrind, still the debugging information 
of the untrusted code was not loaded. This made it difficult to connect the results to the original source code. 
The reason was in the way Valgrind and NaCl load the  binary. First the untrusted code is loaded to a temporary address 
(with \textit{read-only} permission), then NaCl copies the code into the sandboxing area with appropriate permissions 
(code section with \texttt{rx} and data section with \texttt{rw}). This causes Valgrind to lose the association between 
debugging information and the execution addresses. As mentioned before, we fixed the untrusted binary execution location, 
so when Valgrind tries to load the untrusted binary at first, we force it to look for the debugging information, and if present, 
load it and associate it with the execution location of the untrusted binary.
This way we made it possible to execute NaCl under  Valgrind. It can be helpful for debugging purposes as the behavior 
of untrusted code can be monitored inside the sandbox, a capability which was not available before.

\section{Future Work}
\label{future}
Next we enumerate a few open directions for further research.
The greedy algorithm for choosing padding instructions to remove
(Algorithm~\ref{pad_removal}) already appears to achieve a good
proportion of the available overhead reduction, and it is
straightforward to implement because it reuses the validator using its
existing interface.
However repeatedly re-validating portions of object files is
inefficient, and there is no guarantee that the greedy approach will
leave a minimal number of padding bytes overall.
Intuitively, we expect that an efficient algorithm that is close to
optimal in removing padding should be possible by taking a dynamic
programming approach.
The key observation is that for the most part, the validation of an
instruction does not depend on its entire address, only on its
position relative to a bundle boundary, i.e. its address modulo 32.
For instance, there is never a reason to pad by more than 31 bytes.
Thus considering 32 (equivalence classes of) locations for each
instruction is sufficient to choose locations that globally minimize
the need for padding bytes.
Our upper bound measurements of overhead when all type-3 padding is
removed give an idea of the maximum speedup that can be achieved with
better padding removal, though there are also possible layout changes
not covered by that upper bound, such as re-ordering instructions (using semantically re-orderings, or with
added direct jumps). Such changes might also be used to avoid conflicts that disrupt indirect branch prediction.

\section{Conclusion}
\label{concl}
In this paper we investigated the effect of padding on the performance of Software-based Fault Isolation. We took Native Client as one of the most recent and advanced implementations of SFI and changed its padding scheme while still providing the same level of security. Our changes in the NaCl padding scheme demonstrate the possibility of improving SFI performance by avoiding conservative padding. This performance improvement is a result of decrease in the number of instructions executed and better use of instruction caches and prefetch.

Based on our evaluations, because the binaries with bigger code sizes put more pressure on the CPU instruction caches and prefetch, our changes yield in higher performance improvement in such cases. As we reported, the running time reduction for SPECint2006 was on average 8.6\%, while this reduction for SPECint2000 was on average 4.78\%. These results support our intuition about the negative effect of instruction padding on instruction cache and prefetch.

The change in the NaCl padding scheme is implemented in the NaCl assembler. An updated version of the  NaCl validator is provided to cover all the possible instruction streams of the binary which the assembler generates. A formal proof of correctness of this new validator is provided, too. In a detailed performance analysis for x86-32 NaCl (supported by a NaCl-compatible Valgrind tool), we noticed in some cases the new address layout caused by pad removal can lead to higher branch mispredictions. This in turn can cancel out the performance benefits of executing fewer instructions or suffering fewer instruction cache misses.
We also implemented the same modification for x64 NaCl code production
and validation.
We found that few changes were needed to the technique, but the
performance benefit was less than for x86-32 because fewer overlapping
instructions can safely be allowed to execute.

\section*{Acknowledgments}

We thank the anonymous reviewers for suggestions which have helped us
improve the paper's presentation.
The question of whether alignment-based SFI could be improved by
allowing some overlapping instructions was originally suggested by
Greg Morrisett.
This research was supported by the University of Minnesota, and by the
National Science Foundation under grant no. 1526319.

\IEEEtriggeratref{25}


\bibliographystyle{IEEEtranS}
\bibliography{references}
%




\appendix[RockSalt Proof Details]

\label{app:coq-details}

The Coq code excerpts in Figures~\ref{fig:coq-defs} and
\ref{fig:coq-lemma} show the new definition we added to the RockSalt
model, and a key lemma we proved about it.
The recursive function {\tt process\_buffer\allowbreak\_align} corresponds to the
outer loop of the multipass verification algorithm; an existing
function {\tt process\_buffer\_aux} corresponds to the inner loop.
We modified the higher-level function {\tt process\_buffer} to call
the new function.
The first of the sets returned by these functions corresponds to the
bitmap {\em valid} in Figure~\ref{multipass}.
The lemma states that when an instruction address is added to the
{\em valid} set in a successful run of the validator, it must be that
there was a valid instruction at that address.
The proof uses induction over the execution of {\tt
  process\_buffer\_align}, and appeals to a similar lemma about the
inner function {\tt process\_buffer\_aux} as well as a number of
sub-lemmas about arithmetic relationships and list-processing
functions.
The full proof details are available online at \url{https://pastebin.com/gN026Hte} .

\begin{figure*}
\begin{verbatim}
Fixpoint process_buffer_align (loc: int32) (n: nat) (tokens: list token_id)
       (curr_res: Int32Set.t * Int32Set.t) :=
    let (start_instrs, check_list) := curr_res in
      match tokens with
        | nil => Some curr_res
        | _ => (* There are left over bytes in the buffer *)
          match n with
            | O => None
            | S p =>
              match process_buffer_aux loc n tokens (start_instrs, check_list) with
                | None => None
                | Some (start_instrs_new, check_list_new) =>
                   process_buffer_align ( loc +32_n Z.to_nat chunkSize)
                     p (skipn (Z.to_nat chunkSize) tokens)
               (start_instrs_new, check_list_new)
           end
       end
   end.

Definition process_buffer (buffer: list int8) :=
  process_buffer_align (Word.repr 0) (length buffer) (List.map byte2token buffer)
      (Int32Set.empty, Int32Set.empty).
\end{verbatim}
  \caption{Parts of the RockSalt model added or changed to model the
    multipass validator}
\label{fig:coq-defs}
\end{figure*}

\begin{figure*}
\begin{verbatim}
Lemma process_buffer_align_inversion :
 forall n start tokens currStartAddrs currJmpTargets allStartAddrs allJmpTargets,
  process_buffer_align start n tokens (currStartAddrs, currJmpTargets) =
    Some (allStartAddrs, allJmpTargets)
    -> noOverflow (start :: int32_of_nat (length tokens - 1) :: nil)

    -> Z_of_nat (length tokens) <= w32modulus
    -> forall pc:int32, Int32Set.In pc (Int32Set.diff allStartAddrs currStartAddrs)
         -> exists tokens', exists len, exists remaining,
              tokens' = (List.skipn (Zabs_nat (unsigned pc - unsigned start))
                           tokens) /\
              goodDefaultPC_aux (pc +32_n len) start allStartAddrs
                (length tokens) /\
              (dfa_recognize non_cflow_dfa tokens' = Some (len, remaining) \/
               (dfa_recognize dir_cflow_dfa tokens' = Some (len, remaining) /\
                includeAllJmpTargets pc len tokens' allJmpTargets) \/
               dfa_recognize nacljmp_dfa tokens' = Some (len, remaining)).
\end{verbatim}
  \caption{A key new lemma proved about the multipass validator,
    analogous to one about the single-pass validator}
\label{fig:coq-lemma}
\end{figure*}


\end{document}